\documentclass[aps,prb,twocolumn,floatfix,superscriptaddress,showpacs,psfig,reprint, longbilbiography]{revtex4-2}
\usepackage[colorlinks=true,urlcolor=blue,citecolor=blue,linkcolor=blue]{hyperref}
\usepackage{graphicx}
\usepackage{color}
\usepackage{epsfig}
\usepackage{amsmath}
\usepackage{amssymb}
\usepackage{mathrsfs}
\usepackage{bm}
\usepackage{wasysym}
\usepackage{subfigure}
\usepackage{float}
\usepackage{url}
\usepackage{enumerate}
\usepackage{braket}
\usepackage{comment}
\usepackage{multirow}
\usepackage{appendix}
\usepackage[section]{placeins}
\definecolor{violet}{rgb}{0.56,0.0,1.0}

\begin{document}

\hsize\textwidth\columnwidth\hsize\csname@twocolumnfalse\endcsname

\title{Quantum dynamics in a spin-1/2 square lattice $J_{1}$-$J_{2}$-$\delta$ altermagnet}

\author{Yang Liu}
\affiliation{School of Physical Science and Technology $\&$ Key Laboratory of Quantum Theory and Applications of MoE, Lanzhou University, Lanzhou 730000, China.}
\affiliation{Lanzhou Center for Theoretical Physics, Key Laboratory of Theoretical Physics of Gansu Province, Lanzhou University, Lanzhou 730000, China.}
\author{Shiqi Shao}
\affiliation{School of Physical Science and Technology $\&$ Key Laboratory of Quantum Theory and Applications of MoE, Lanzhou University, Lanzhou 730000, China.}
\affiliation{Lanzhou Center for Theoretical Physics, Key Laboratory of Theoretical Physics of Gansu Province, Lanzhou University, Lanzhou 730000, China.}
\author{Saisai He}
\affiliation{School of Physical Science and Technology $\&$ Key Laboratory of Quantum Theory and Applications of MoE, Lanzhou University, Lanzhou 730000, China.}
\affiliation{Lanzhou Center for Theoretical Physics, Key Laboratory of Theoretical Physics of Gansu Province, Lanzhou University, Lanzhou 730000, China.}
\author{Z. Y. Xie}
\email{qingtaoxie@ruc.edu.cn}
\affiliation{School of Physics, Renmin University of China, Beijing 100872, China}
\affiliation{Key Laboratory of Quantum State Construction and Manipulation of MoE, Renmin University of China, Beijing 100872, China}
\author{Jia-Wei Mei}
\affiliation{Department of Physics, Southern University of Science and Technology, Shenzhen 518055, China}
\author{Hong-Gang Luo}
\affiliation{School of Physical Science and Technology $\&$ Key Laboratory of Quantum Theory and Applications of MoE, Lanzhou University, Lanzhou 730000, China.}
\affiliation{Lanzhou Center for Theoretical Physics, Key Laboratory of Theoretical Physics of Gansu Province, Lanzhou University, Lanzhou 730000, China.}
\author{Jize Zhao}
\email{zhaojz@lzu.edu.cn}
\affiliation{School of Physical Science and Technology $\&$ Key Laboratory of Quantum Theory and Applications of MoE, Lanzhou University, Lanzhou 730000, China.}
\affiliation{Lanzhou Center for Theoretical Physics, Key Laboratory of Theoretical Physics of Gansu Province, Lanzhou University, Lanzhou 730000, China.}

\begin{abstract}
A key feature of the newly discovered altermagnet is that its spin degeneracy is lifted, although it has an antiferromagnetic order and zero net magnetization.  
In this work, we investigate a frustrated spin-1/2 $J_1$-$J_2$-$\delta$ Heisenberg model on the square lattice by the tensor network method 
in combination with the linear spin-wave theory, with our focus on both the magnon excitations and longitudinal excitations. 
For a small $J_2$ and a finite range of $\delta$ we demonstrate that such a model hosts an altermagnetic ground state. 
Its magnon spectrum is split into two branches and the largest splitting occurs at $\left(\pm\pi/2, \pm\pi/2\right)$ in the Brillouin zone.
The magnitudes of splitting in the two magnon modes are equal with respect to the case of $\delta=0$. 
Dynamical spin structure factors show that the low-energy peak in the longitudinal spectral weight around $(\pi/2, \pi/2)$ is also split, 
and thus the relative positions of the magnon modes and longitudinal modes in energy may change in the presence of a finite $\delta$. 
These findings demonstrate that the altermagnets harbor more complex quantum dynamics than the conventional collinear antiferromagnets.
\end{abstract}

\pacs{}
\maketitle
\textit{Introduction.} 
Very recently, a series of studies~\cite{PhysRevB.99.184432, doi:10.7566/JPSJ.88.123702, PhysRevB.102.014422, PhysRevB.102.144441, doi:10.1073/pnas.2108924118, doi:10.1126/sciadv.aaz8809, Ma2021, PhysRevX.12.040002} have led to the discovery of a new type of collinear magnets dubbed altermagnets~\cite{PhysRevX.12.031042, PhysRevX.12.040501}, 
which are distinct from the well-known collinear ferromagnets and antiferromagnets. In the altermagnets, nearest-neighbor spins are aligned antiparallel, 
resembling those in the antiferromagnets. However, due to the inequivalent environments around the two nearest-neighbor spins, the spin degeneracy of the energy band is lifted 
with alternating sign in the Brillouin zone~\cite{PhysRevX.12.031042, PhysRevX.12.040501, Krempasky2024}. 
Thus the altermagnets share some properties with ferromagnets, though they maintain zero net magnetization. This alternating spin splitting may lead to some exotic properties such as anomalous transport behaviors~\cite{https://doi.org/10.1002/adfm.202409327, PhysRevB.108.L180401, PhysRevLett.132.086701, PhysRevLett.133.086503} which are absent in the conventional antiferromagnets. Therefore, altermagnets are argued to be a promising candidate for future spintronics and magnonics. 

Theoretically, to gain an intuitive understanding of correlated systems , excitations are usually approximated as quasiparticles
based on their degrees of freedom and symmetry. For example, in altermagnets as well as in other magnetically ordered systems, 
the linear spin-wave theory~(LSWT) demonstrates that the low-energy excitations are the quanta of spin waves, and the quasiparticles are called magnons. 
A straightforward observation in altermagnets is that the magnon modes with opposite chiralities are split, resulting in chiral magnons~\cite{PhysRevLett.131.256703}. 
However, this simplified picture may be challenged when high-energy excitations become significant~\cite{Merchant2014, Hong2017, PhysRevLett.122.127201}. 
In particular, LSWT is limited to magnon excitations and other excitations are completely neglected. 
Moreover, the approximation in LSWT is justified only in the large $S$ limit where the quantum fluctuations are negligible. 
For spin-1/2 systems, it usually fails in reproducing experimental results~\cite{PhysRevB.71.184440, Syromyatnikov_2010} quantitatively.
Moreover, it neglects magnon-magnon interactions and other types of interactions, which are known to become increasingly important at high energies in correlated systems.
To gain a comprehensive understanding of the dynamical properties of an altermagnet with strong quantum fluctuations, 
we resort to the new development in the tensor network states~(TNS) method~\cite{PhysRevB.92.201111, PhysRevB.101.195109, 10.21468/SciPostPhys.12.1.006} to investigate the 
spin-1/2 $J_1$-$J_2$-$\delta$ Heisenberg model on the square lattice. 

\begin{figure}[ht]
        \includegraphics[width=1.0\columnwidth, clip]{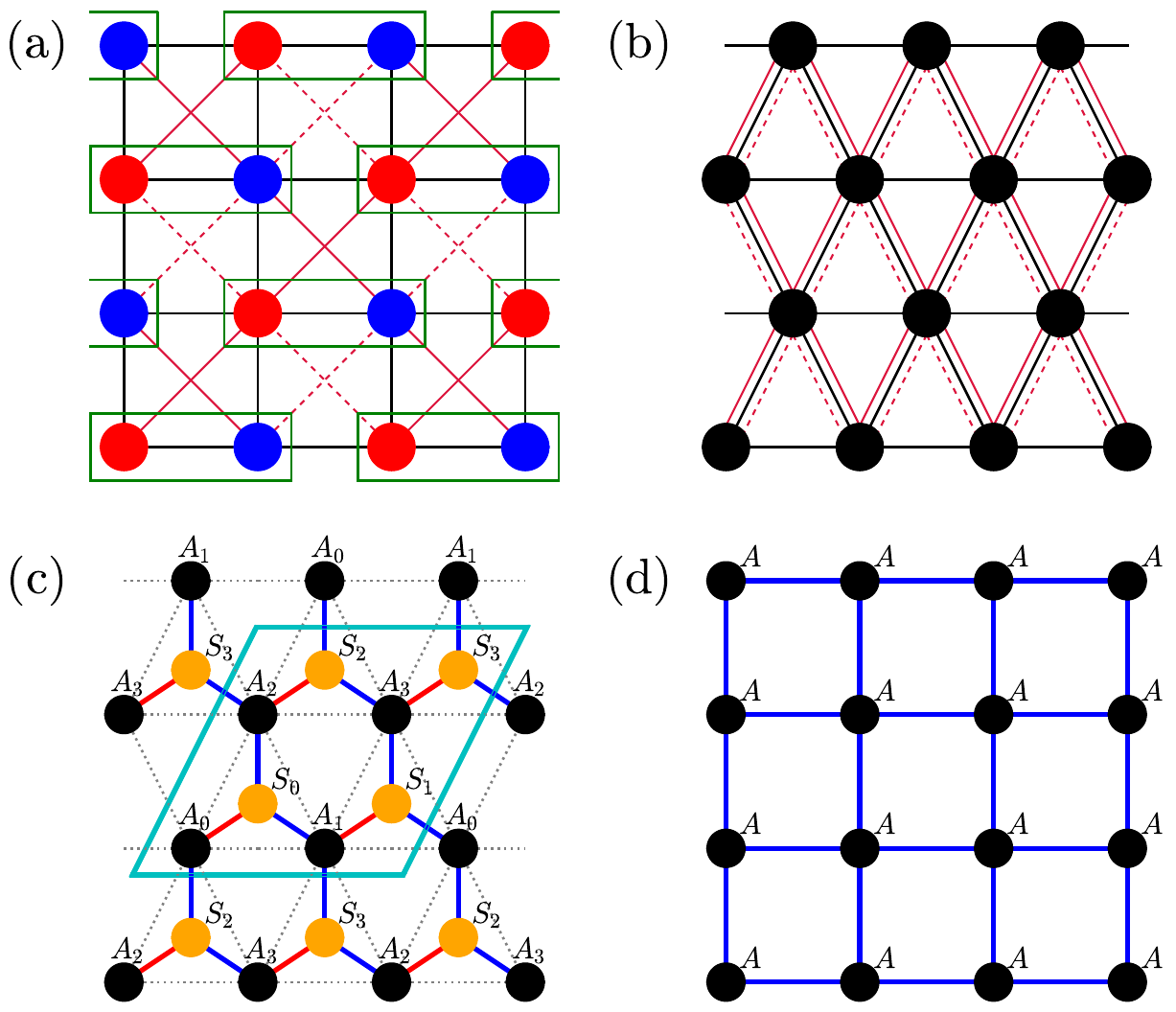}
	\caption{(a) A schematic picture illustrating the $J_1$-$J_2$-$\delta$ model on the square lattice. Filled circles represent the spins. 
	Two types of inequivalent spins are marked in red and blue, respectively. The black solid lines plot the nearest-neighbor interaction $J_1$. 
	The red solid lines and dashed lines are the next nearest-neighbor interactions with the couplings $J_2(1+\delta)$ and $J_2(1-\delta)$, respectively. 
	(b) The triangular lattice obtained from (a) by merging two spins within the green rectangle into one site~(black circle). (c) The iPESS ansatz used (with a $2\times 2$ 
	unit cell) in the imaginary-time evolution, obtained by introducing a simplex tensor~(orange circle) in each upper triangle in (b). 
	(d) The iPEPS ansatz used in the variational calculation, obtained by merging the simplex tensor (orange) and the projection tensor (black) connected by the red lines in (c) together.}
       \label{lattice}
\end{figure}

\textit{Model and altermagnetic ground state.}
The spin-1/2 $J_{1}$-$J_{2}$-$\delta$ Heisenberg model, as plotted in Fig.\ref{lattice}(a), is described by following Hamiltonian,
\begin{equation}
        \mathcal{H} = J_1\sum_{\langle{i,j}\rangle}{\bf{S}}_{i}\cdot{\bf{S}}_{j}+J_{2}\sum_{\langle\langle{i,j}\rangle\rangle}(1\pm\delta)\mathbf{S}_{i}\cdot\mathbf{S}_{j},\label{HAM}
\end{equation}
where $J_1(>0)$ is the nearest-neighbor coupling and set as the energy unit. $J_2(>0)$, modulated by the factor $1\pm\delta$, is the next nearest-neighbor coupling. 
Here $\pm\delta$ mimics the inequivalent surrounds of the nearest-neighbor spins and the checkerboard lattice corresponds to $\delta=1$. 
Recently, P. Das {\it{et al.}} proposed a scheme to realize altermagnetism 
with ultracold fermion atoms in the optical lattice~\cite{PhysRevLett.132.263402}. Model~(\ref{HAM}) can be realized by their scheme~\cite{PhysRevLett.132.263402} 
with a large onsite interaction and in iron oxychalcogenides~\cite{PhysRevLett.104.216405}.   
In addition, monolayer $\mathrm{V_2Se_2O}$ and $\mathrm{V_2Te_2O}$ may possibly be modelled by its spin-1 counterpart~\cite{Ma2021}.  

When $\delta=0$, Model~(\ref{HAM}) is reduced to the spin-1/2 $J_1$-$J_2$ Heisenberg model. Previous works have consistently demonstrated that such a model
hosts a conventional collinear antiferromagnet when $J_2\apprle{0.4J_1}$. We expect that for a small $J_2$ such an antiferromagnetic order may 
extend to a finite $\delta$, thus favoring the altermagnetism. Before we dedicate ourselves to studying the excitation spectrum, it is necessary to conduct systematic TNS 
simulations~\cite{RevModPhys.93.045003, PhysRevLett.101.090603, doi:10.1143/JPSJ.65.891, PhysRevLett.113.046402, PhysRevB.82.245119, XFLiu67502} to confirm this conjecture. For this purpose, as illustrated in Fig.~\ref{lattice}(a) and (b), we first deform the square lattice into a triangular lattice by merging the two sites within the green rectangle, and then arrange a simplex tensor in the center of each upper triangle~\cite{PhysRevB.105.184418, PhysRevB.107.L041106}, as shown in Fig.~\ref{lattice}(c). 
This is the infinite projected entangled simplex state~(iPESS) ansatz~\cite{PhysRevX.4.011025}, which is known to be efficient for frustrated lattice models. 
The ansatz may be optimized by the imaginary-time evolution equipped with the simple-update~(SU) strategy~\cite{PhysRevLett.101.090603}. 
To accelerate the simulation, a local spin rotation $-i\sigma_y$ is performed~\cite{10.21468/SciPostPhys.10.1.012} on the operators at the blue sites in Fig.~\ref{lattice}(a).

\begin{figure}[!ht]
	\centering
	\includegraphics[width=1.0\columnwidth, clip]{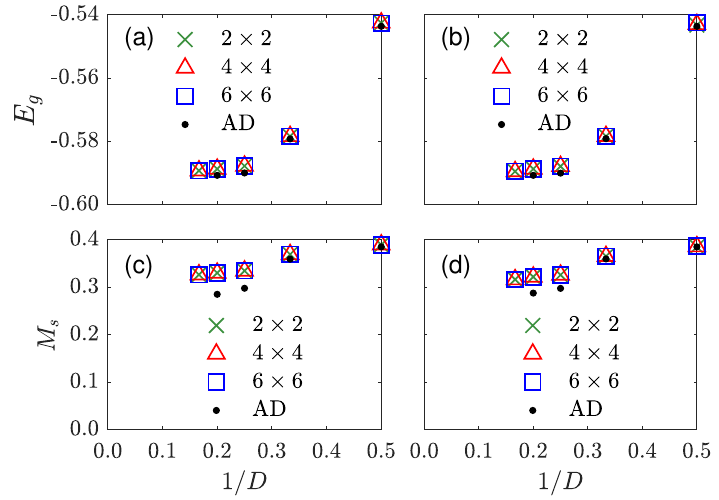}
	\caption{(a)-(b) The ground-state energy per site $E_g$ and (c)-(d) the single site magnetization $M_s$ are shown as a function of $1/D$. 
        The model parameters are $J_2=0.2$, $\delta=0.5$~(left) and $\delta=1.0$~(right).
	Three unit cells $2\times{2}$, $4\times{4}$ and $6\times{6}$ in PESS ansatz are used in SU. 
	The filled black circles represent the AD results.}
	\label{gs}
\end{figure}

In this work, we have conducted SU calculations on three different iPESS ansatz with unit cell size $2\times2$, $4\times4$ and $6\times6$. 
The results, presented in Fig.~\ref{gs}, are shown as a function of $1/D$ with $D$ the bond dimension of the ansatz. 
We can see that $E_g$ and $M_s$ are nearly identical for three different unit cells and they converge quickly as $D$ increases, 
demonstrating the stability of our iPESS ansatz for the given parameters. Moreover, we observed that $M_s$ is site-independent and nearest-neighbor spins are aligned antiparallel,
strongly supporting our assertion that the ground state is altermagnetic.

To obtain the physical quantities with a relatively high accuracy, we further deform the iPESS ansatz with a $1\times 1$ unit cell, which is the simplest one 
compatible with the ground state configuration, into a single-site infinite projected entangled pair state~(iPEPS)~\cite{2004cond.mat7066V}, as shown in Fig.~\ref{lattice}(d). 
Then the wave function is optimized variationally by the automatic differentiation (AD)~\cite{PhysRevX.9.031041, PhysRevB.101.220409} technique. As shown in Fig.~\ref{gs}, 
the obtained ground state is also altermagnetic but with a more accurate energy. 
In the following, our main results are obtained using the AD technique with $D=4$, unless stated explicitly.

\textit{Magnon dispersion.}
\begin{figure*}[htbp]
        \includegraphics[width=0.98\textwidth, clip]{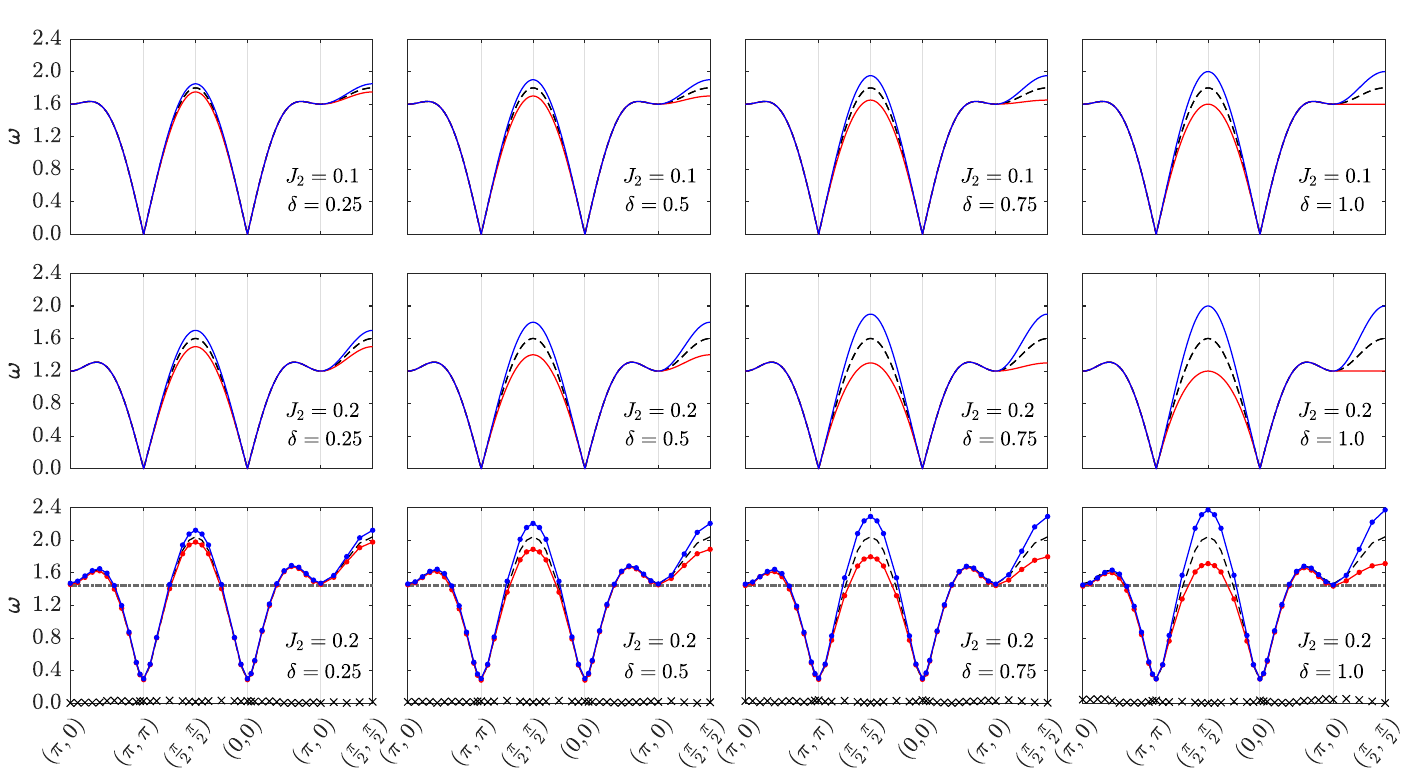}
	\caption{The magnon spectrum are shown for $\delta=0.25, 0.50, 0.75, 1.0$~(solid lines) and for $\delta=0$~(dashed lines). 
	The first two rows are obtained by the LSWT and the last row 
	is obtained by the TNS. In the last row the dot-dashed lines at $\omega=1.4456$ mark the height of the roton-like minimums
	and the crosses~($\times$) mark $|\delta\omega^{+}(\mathbf{k})-\delta\omega^{-}(\mathbf{k})|$ 
	with $\delta\omega^{+}(\mathbf{k})=\omega^+(\mathbf{k})-\omega^0(\mathbf{k})$ and $\delta\omega^{-}(\mathbf{k})=\omega^0(\mathbf{k})-\omega^-({\mathbf{k}})$.} 
        \label{magnon}
\end{figure*}
Now we turn to presenting the magnon dispersion obtained by the LSWT and TNS methods, respectively. 
In the $J_1$-$J_2$-$\delta$ model, the LSWT~\cite{Toth_2015} gives two branches of magnon modes,
\begin{equation}
	\omega^{\pm}({\bf{k}}) = \frac{\sqrt{c_2^2-4c_1^2}}{2}\pm{\frac{c_3}{2}}, \label{LSDS} 	
\end{equation}
where $c_1=J_1\left(\cos(k_x)+\cos(k_y)\right)$, $c_2=4J_1-4J_2(1-\cos(k_x)\cos(k_y))$, and $c_3=4J_2\delta\sin(k_x)\sin(k_y)$. When $c_3$ is nonzero, the 
magnon spectrum is split into two branches $\omega^+({\bf{k}})$ and $\omega^-({\bf{k}})$. The maximum splitting occurs at $(\pm\pi/2, \pm\pi/2)$. 
On the other hand, the spectrum is degenerate when $c_3=0$, i.e., along the line $(0, k_y)$, $(\pi, k_y)$, $(k_x, 0)$ and $(k_x, \pi)$. 
This reflects the sign of splitting is alternating in the Brilliuin zone, which is another characteristic feature of the altermagnets. 
We notice that only $c_3$ depends on $\delta$, which tells us that the magnon spectrum $\omega^\pm(\mathbf{k})$ is equally split with respect to 
the magnon dispersion $\omega^0(\mathbf{k})$ for $\delta=0$.
In Fig.~\ref{magnon}, we plot the magnon excitations along the path 
$(\pi,0)-(\pi,\pi)-(\frac{\pi}{2},\frac{\pi}{2})-(0,0)-(\pi,0)-(\frac{\pi}{2},\frac{\pi}{2})$. 
The first two rows, where $J_2=0.1$ and 0.2, are obtained by the LSWT. 
This figure clearly shows the characteristic features of the altermagnets stated above. Moreover, 
both branches exhibit a roton-like minimum at $(\pi,0)$. Such an excitation was first introduced by Landau to explain 
the superfluidity in helium. In the literature, several scenarios have been proposed to account for the roton-like minimum~\cite{PhysRevLett.115.207202, DallaPiazza2015, PhysRevX.7.041072, 10.21468/SciPostPhys.4.1.001, PhysRevB.98.100405, PhysRevB.98.134410, PhysRevB.98.155102}.
Here we simply attribute the reason to the fact that $J_2$ favors a stripe order with the wavevector $\mathbf{k}=(\pi,0)$ or $(0,\pi)$. 
The roton-like minimum is doubly degenerate with $\omega^{+}(\pi,0)=\omega^{-}(\pi,0)=2J_1-4J_2$, 
which is independent of the parameter $\delta$. Moreover, it softens as $J_2$ increases, indicating a trend to the stripe phase.  
One may notice that the band along the line $(\pi,0)$-$(\pi/2, \pi/2)$ is flattened as $\delta$ increases, which becomes dispersionless at $\delta=1.0$. 
As $\delta$ increases further, the roton-like minimum in $\omega^-(\mathbf{k})$ disappears, and instead, a saddle point appears at $(\frac{\pi}{2}, \frac{\pi}{2})$.

In the third row we show the excitation spectrum obtained by the TNS method,
which is quantitatively convincing at intermediate and high energies~\cite{PhysRevB.101.195109, 10.21468/SciPostPhys.12.1.006, PhysRevLett.129.227201}. 
To be specific, we fix $J_2=0.2$. Although the magnon spectrum are strongly renormalized we do observe some qualitatively consistent results in comparison with the predictions of LSWT.
The splitting of $\omega^\pm(\mathbf{k})$ is clearly seen along the paths from $(\pi,\pi)$ to $(0,0)$
and from $(\pi,0)$ to $(\pi/2, \pi/2)$ with the maximum splitting at $(\pi/2, \pi/2)$. 
$\omega^0(\mathbf{k})$ sits between $\omega^+(\mathbf{k})$ and $\omega^-(\mathbf{k})$, and it is roughtly equal 
to $\left(\omega^+(\mathbf{k})+\omega^-(\mathbf{k})\right)/2$. A roton-like minimum is found at $(\pi,0)$ and within the numerical error the roton gap is independent of $\delta$.
Such a gap softens as $J_2$ increases~(see Supplemental Material~(SM)~\cite{Supp}. These conclusions agree well with the predictions of LSWT.
As $\delta$ increases, the magnon band along $(\pi,0)$-$(\pi/2, \pi/2)$ becomes more flat. However, at $\delta=1$ it remains dispersive, 
which is different from the conclusion given by Eq.~(\ref{LSDS}). Higher order corrections
to LSWT should be taken into account to explain such difference. 
One may notice that magnon spectrum at $(\pi,\pi)$ is gapped and it seems that TNS cannot reproduce the Nambu-Goldstone mode. 
Actually this is the finite $D$ effect~\cite{PhysRevB.101.195109, 10.21468/SciPostPhys.12.1.006, PhysRevLett.129.227201}. In the SM~\cite{Supp}, 
we demonstrate that such a gap decreases as $D$ increases and eventually disappears in the large $D$ limit.    
\begin{figure*}[htb]
        \includegraphics[width=1.0\textwidth, clip]{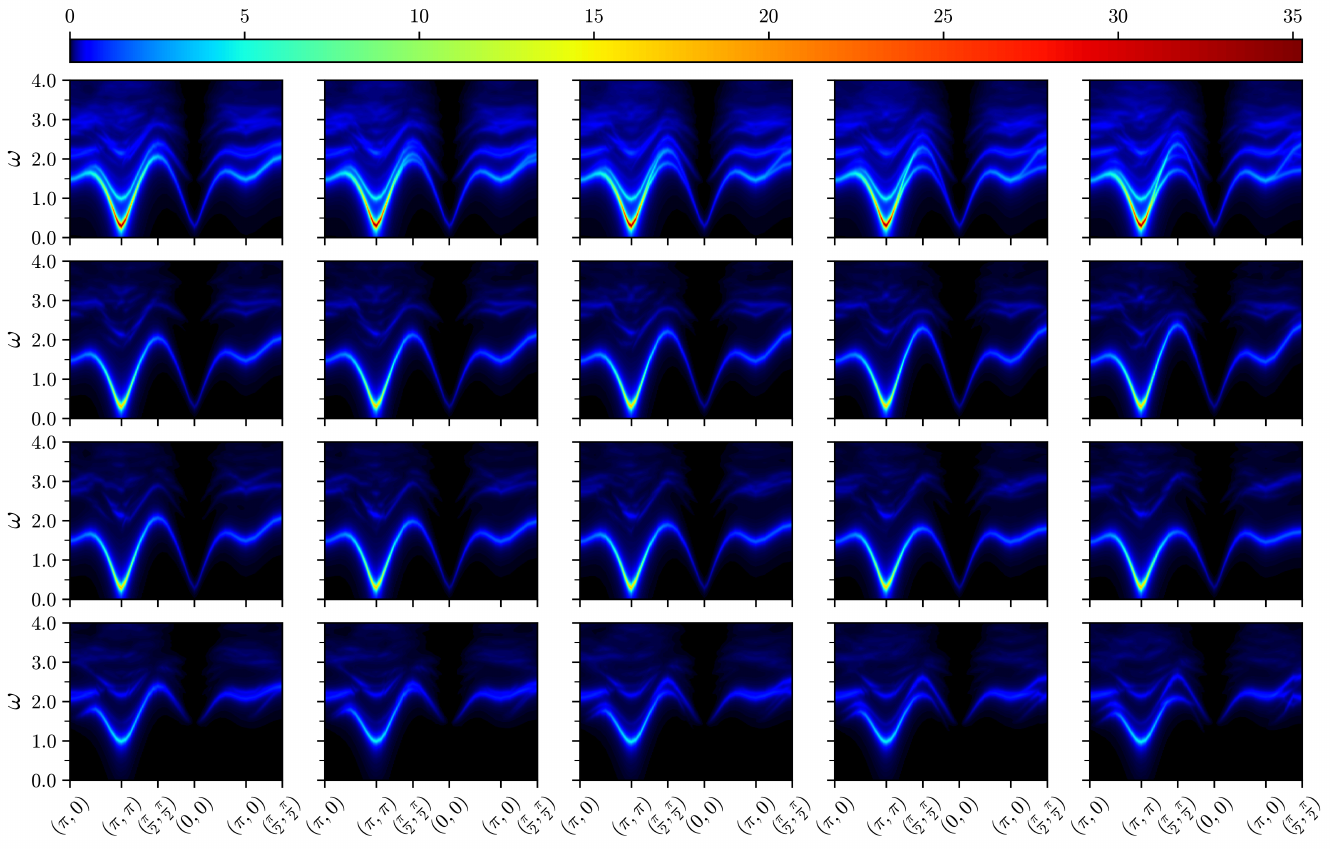}
        \caption{Dynamical spin structure factors are shown along high-symmetry lines. $J_2=0.2$ and $\delta$ takes 0.0, 0.25, 0.5, 0.75 and 1.0 from left to right.
	The rows from top to bottom correspond to $S(\mathbf{k},\omega)$,
        $S^{-+}(\mathbf{k}, \omega)/2$, $S^{+-}(\mathbf{k},\omega)/2$ and $S^{l}(\mathbf{k}, \omega)$. Note that when $\delta$ is finite $S^{-+}(\mathbf{k}, \omega)$
        and $S^{+-}(\mathbf{k}, \omega)$ are different. For the given ground state, the dominant contributions to $S^{-+}(\mathbf{k}, \omega)$ and $S^{+-}(\mathbf{k}, \omega)$
        are from the $\omega^+(\mathbf{k})$ and $\omega^-(\mathbf{k})$ magnon modes, respectively. In the $S^{l}(\mathbf{k}, \omega)$ there are sharp 
	longitudinal modes.}
        \label{FIG4}
\end{figure*}

\textit{Dynamic spin structure factors.}
To further analyze the nature of the excitations and provide references for experiments, we calculate the dynamical spin structure factors~(DSSFs) 
which are directly measurable in the inelastic neutron scattering experiments. The DSSFs is defined as 
\begin{equation}
\begin{split}
	S^{\alpha\beta}(\mathbf{k},\omega) & = \sum_{n>0}\langle0|S^{\alpha}_{-\bold{k}}|n\rangle\langle{n}|S^{\beta}_{\bold{k}}|0\rangle\delta(\omega-\omega_n).
\end{split}
\end{equation}
where $S_\mathbf{k}^\alpha=\sum_i \exp(i\mathbf{k}\cdot\mathbf{R}_i)S^\alpha_i$ and $\alpha, \beta=z$ or $+,-$. Here for simplicity we transform the polarization 
direction to the z direction. The $\delta$ function is approximated by a Lorentzian distribution with a broadening of 0.06$J_1$ to mimic the influence of the environment in experiments. 
In the collinear magnetic order, we are interested in the longitudinal part $S^l(\mathbf{k},\omega)=S^{zz}(\mathbf{k},\omega)$,  
the transverse part $S^t(\mathbf{k},\omega)=\left(S^{+-}(\mathbf{k},\omega)+S^{-+}(\mathbf{k}, \omega)\right)/2$ and their summation $S(\mathbf{k},\omega)$. 

\begin{figure}[!hbt]
   \includegraphics[width=0.8\columnwidth]{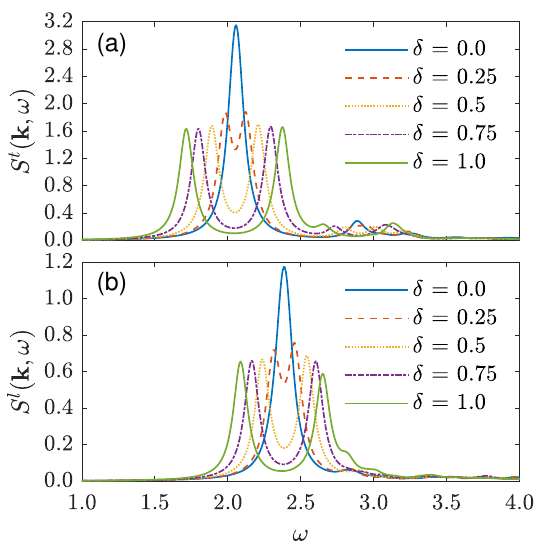}	
	\caption{The spectral weight $S^t(\mathbf{k},\omega)$ and $S^l(\mathbf{k},\omega)$ at $\mathbf{k}=(\pi/2,\pi/2)$ are shown in (a) and (b), respectively.}
	\label{FIG5}
\end{figure}

In Fig.~\ref{FIG4}, we plot the DSSFs for $\delta=0.0, 0.25, 0.5, 0.75$ and 1.0, respectively. In addition to the sharp magnon modes, 
we observe sharp longitudinal modes~(fourth row) in $S(\mathbf{k},\omega)$~\cite{Tsuji_2024,Pekker_2015}, 
on top of which the high-energy part of the spectrum is occupied by the continuum. 
As shown in the second and third rows of Fig.~\ref{FIG4}, in the conventional collinear antiferromagnet, i.e., $\delta=0$,
we have $S^{-+}(\mathbf{k},\omega) = S^{+-}(\mathbf{k}, \omega)$. 
However, for a finite $\delta$ such degeneracy is lifted, leading to the conclusion $S^{-+}(\mathbf{k},\omega)\ne{S^{+-}(\mathbf{k}, \omega)}$. 
This difference may serve as a experimental signal to distinguish the antiferromagnet and altermagnet.
For all $\delta$, the dominant spectral weight in $S(\mathbf{k},\omega)$ appears near $\mathbf{k}=(\pi,\pi)$ while it is vanishingly small near the $\mathbf{k}=(0,0)$ point.
This reflects the antiparallel ground-state spin configurations. Moreover, we find that at $\mathbf{k}=(\pi,\pi)$, the intensity of $S(\mathbf{k},\omega)$ is almost independent of 
$\delta$ and thus its behavior has been extensively discussed. We refer the reader to Ref.~\cite{PhysRevLett.115.207202, DallaPiazza2015, PhysRevX.7.041072, 10.21468/SciPostPhys.4.1.001, PhysRevB.98.100405, PhysRevB.98.134410, PhysRevB.98.155102, PhysRevB.99.224402} for details. 
Significant difference is observed around $\mathbf{k}=(\pi/2, \pi/2)$, where the energy splitting in magnon modes are most apparent. Moreover, 
the low-energy peak in $S^l(\mathbf{k}, \omega)$ is also split (see also Fig.~\ref{FIG5}), indicating the relative positions of the magnon modes and longitudinal modes may change. In particular, when $\delta\apprge{0.5}$ the lower longitudinal modes in energy are lower than the upper magnon mode and 
thus may become more stable~\cite{PhysRevB.102.125102}. These may be detected by polarized neutron experiments with high resolutions. 

In Fig.~\ref{FIG5} we show the spectral weights $S^t(\mathbf{k},\omega)$ and $S^l(\mathbf{k},\omega)$ at $\mathbf{k}=(\pi/2,\pi/2)$ for a variety of $\delta$. 
It is clearly demonstrated that the low-energy peaks are split into two branches in the presence of a finite $\delta$ and
thereafter the spectral weights evolve continuously as $\delta$ increases. Here we want to emphasize that the splitting in $S^l(\mathbf{k}, \omega)$ 
may result from the repelling of energy levels and it does not necessarily indicate the lift of degeneracy.  
The peak height of the two magnon modes (Fig.~\ref{FIG5}(a)) and 
lower longitudinal mode (Fig.~\ref{FIG5}(b)) approaches a constant 
and no obvious broadening is observed as $\delta$ increases, demonstrating the stability of these excitation modes.  
However, the upper longitudinal mode decreases monotonically and tend to merge with the continuum.  

\textit{Conclusion.}
Quantum dynamics of the spin-1/2 $J_{1}$-$J_{2}$-$\delta$ Heisenberg model on the square lattice is studied using the TNS method and LSWT. 
When $J_2$ is small our TNS simulations confirm that the antiparallel spin configuration extends to a finite range of $\delta$, thus favoring altermagnetism.
The dynamical spin structure factors demonstrate that there are sharp magnon modes and longitudinal modes in the excitation spectrum.
The splitting in the energy spectrum is found in the magnon modes. At some $\mathbf{k}$ points the low-energy longitudinal modes may appear below the magnon mode.  
At $(\pi,0)$ or equivalently $(0,\pi)$ a roton-like minimum is revealed in the magnon spectrum. The roton gap is independent of $\delta$ but softens as $J_2$ increases.
When $\delta\ge{1}$ TNS results show that the roton-like minimum in the lower magnon branch remains stable but LSWT gives an opposite prediction. 
High-order corrections to the LSWT may be necessary to remedy this contradiction. 
We hope that the complex dynamics in the altermagnet are detectable in future inelastic neutron scattering experiments.

\textit{Acknowledgements.}
We are grateful to H.-J. Liao, W. Li and R. Yu for helpful discussions. This work is supported by the National key R$\&$D Program of China (Grants No. 2022YFA1402704, 2023YFA1406500, 2021YFA1400400), by the National Natural Science Foundation of China (Grants Nos. 12274187, 12274458, 12474143, 12247101) and by the Shenzhen Fundamental Research Program (Grant No. JCYJ20220818100405013 and JCYJ20230807093204010).  

\bibliography{AM}

\clearpage

\onecolumngrid

\newpage

\newcounter{equationSM}
\newcounter{figureSM}
\newcounter{tableSM}
\stepcounter{equationSM}
\setcounter{equation}{0}
\setcounter{figure}{0}
\setcounter{table}{0}
\setcounter{page}{1}
\makeatletter
\renewcommand{\theequation}{\textsc{sm}-\arabic{equation}}
\renewcommand{\thefigure}{\textsc{sm}-\arabic{figure}}
\renewcommand{\thetable}{\textsc{sm}-\arabic{table}}


\begin{center}
{\large{\bf Supplemental Material for\\
 ``Quantum dynamics in a spin-1/2 square lattice $J_{1}$-$J_{2}$-$\delta$ altermagnet''}}
\end{center}
\begin{center}
	Yang Liu$^{1,2}$, Shiqi Shao$^{1,2}$, Saisai He$^{1,2}$, Z.Y.Xie$^{3,4}$, Jia-Wei Mei$^5$, Hong-Gang Luo$^{1,2}$, and Jize Zhao$^{1,2}$\\
\quad\\
$^1$\textit{School of Physical Science and Technology $\&$ Key Laboratory of Quantum Theory and Applications of MoE, Lanzhou University, Lanzhou 730000, China.}\\
$^2$\textit{Lanzhou Center for Theoretical Physics, Key Laboratory of Theoretical Physics of Gansu Province, Lanzhou University, Lanzhou 730000, China.}\\
$^3$\textit{School of Physics, Renmin University of China, Beijing 100872, China}\\
$^4$\textit{Key Laboratory of Quantum State Construction and Manipulation of MoE, Renmin University of China, Beijing 100872, China}\\
$^5$\textit{Department of Physics, Southern University of Science and Technology, Shenzhen 518055, China}
\end{center}

\twocolumngrid



\begin{figure}[!ht]
        \includegraphics[width=1.0\columnwidth, clip]{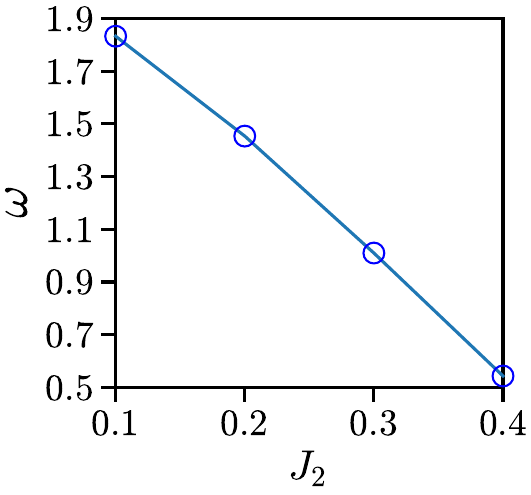}
        \caption{The magnon gap at $(\pi, 0)$, i.e., roton gap, is shown as a function of $J_2$. $\delta=0.5$ is fixed in the Hamiltonian. The solid line is the eye guide.}
        \label{FIGS1}
\end{figure}

$J_2$ term in the $J_1$-$J_2$-$\delta$ model favors a stripe order with a wave vector $(\pi, 0)$ or equivalently $(0, \pi)$.
We expect that as $J_2$ increases the tendency to the stripe order will become stronger, and thus the roton gap will decrease.
To confirm this we calculate the roton gaps by the tensor-networks method. Here $\delta$ is fixed as $\delta=0.5$.
In the tensor-networks calculation, we take the bond dimension $D=4$. Our results are shown as a function of $J_2$ in Fig.~\ref{FIGS1}(a).
This figure clearly shows that the roton gap decreases monotonically as $J_2$ increases.
In this work, we focus on the altermagnetic phase and have no intention to discuss other phases so we stop at $J_2=0.4$.

\begin{figure}[!ht]
        \includegraphics[width=1.0\columnwidth, clip]{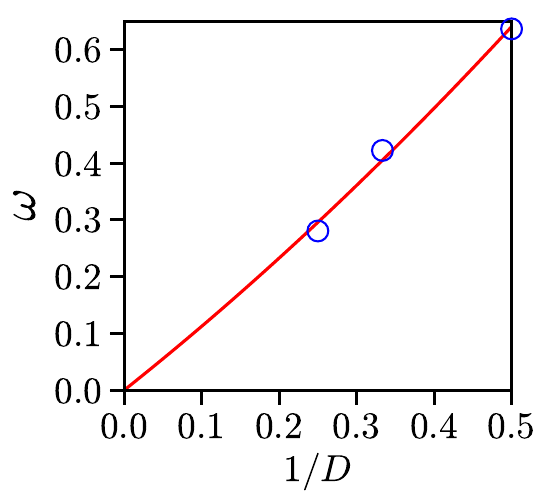}
        \caption{The magnon gap at $(\pi,\pi)$ is plotted as a function of $1/D$. The data are fitted by $\omega=a_0+a_1/D+a_2/D^2$ with $a_0\ge{0.0}$,
        which is shown by the red solid line. $\delta=0.0$ is chosen.}
        \label{FIGS2}
\end{figure}

In the altermagnetic phase, SU(2) symmetry is spontaneously broken. The magnon excitation at $(\pi,\pi)$ should be gapless, which is the so-called Nambu-Goldstone mode.
This is correctly captured by the linear spin-wave theory. However, it seems that the tensor-networks method fails to do so. We will demonstrate that
this is due to the finite $D$ effect. In the large $D$ limit, we can reproduce the Nambu-Goldstone mode. In Fig.~\ref{FIGS2}, we show the
magnon gap at $(\pi,\pi)$ as a function of $1/D$. As $D$ increases the magnon gap decreases and eventually vanishes in the large $D$ limit.

\end{document}